Annals of
**Pure and Applied Mathematics**

# Irregular Interval–Valued Fuzzy Graphs

### *Madhumangal Pal*[1] **and** *Hossein Rashmanlou*[2]

[1]Department of Applied Mathematics with Oceanology and Computer Programming
Vidyasagar University, Midnapore-721102, India
Email: mmpalvu@gmail.com

[2]Department of Mathematics, University of Mazandaran, Babolsar, Iran
Email: hrashmanlou@yahoo.com

**Abstract.** In this paper, we define irregular interval-valued fuzzy graphs and their various classifications. Size of regular interval-valued fuzzy graphs is derived. The relation between highly and neighbourly irregular interval-valued fuzzy graphs are established. Some basic theorems related to the stated graphs have also been presented.

**1. Introduction**
Presently, science and technology is featured with complex processes and phenomena for which complete information is not always available. For such cases, mathematical models are developed to handle various types of systems containing elements of uncertainty. A large number of these models are based on an extension of the ordinary set theory, namely, fuzzy sets. In 1965, Zadeh [31] introduced the notion of fuzzy subset of a set as a method of presenting uncertainty. The fuzzy systems have been used with success in last years, in problems that involve the approximate reasoning. It has become a vast research area in different disciplines including medical and life science, management sciences, social sciences, engineering, statistics, graph theory, artificial intelligence, signal processing, multi agent systems, pattern recognition, robotics, computer networks, expert systems, decision making, etc.

In 1975, Rosenfeld [19] introduced the concept of fuzzy graphs. In 1975, Zadeh [30] introduced the notion of interval-valued fuzzy sets as an extension of fuzzy sets [31] in which the values of the membership degrees are intervals of numbers instead of real numbers between 0 and 1. Interval-valued fuzzy sets provide a more adequate description of uncertainty than traditional fuzzy sets. It is therefore important to use interval-valued fuzzy sets in applications, such as fuzzy control. One of the computationally most intensive parts of fuzzy control is defuzzification [10]. Since interval-valued fuzzy sets are widely studied and used, we describe briefly the work of Gorzalczany on approximate reasoning [8-9], Roy and Biswas on medical diagnosis [20], Turksen on multivalued logic [25] and Mendel on intelligent control [10]. In 2011, Akram [1] introduced the concept of interval-valued fuzzy graphs and defined different operations on it.

Interval- valued fuzzy graph theory is now growing and expanding its applications. The theoretical development in this area is discussed here.

## 1.1. Review of literature

After Rosenfeld [19], fuzzy graph theory is increased with a large number of branches. Mathew and Sunitha [11] described the types of arcs in a fuzzy graph. Nagoorgani and Malarvizhi [14] established the isomorphism properties of strong fuzzy graphs. Nagoorgani and Vadivel [15] established relations between the parameters of independent domination and irredundance in fuzzy graphs. Nair and Cheng [17] defined cliques and fuzzy cliques in fuzzy graphs. Nair [18] established the definition of perfect and precisely perfect fuzzy graphs. Akram [2] defined different operations on bipolar fuzzy graphs. Strong bipolar fuzzy graphs were also introduced here. He also introduced regular bipolar fuzzy graphs [4]. Samanta and Pal introduced fuzzy tolerance graphs [21], fuzzy threshold graphs [23], fuzzy competition graphs [22] and bipolar fuzzy hypergraph [24]. Akram and Davvaz discussed the properties of strong intuitionistic fuzzy graphs [3]. Talebi and Rashmanlou [26] studied isomorphism on interval-valued fuzzy graph. Likewise, they defined isomorphism on vague graphs [27]. A very few algorithms have also been designed to solve problems on fuzzy graphs [32,33,34].

## 2. Preliminaries

A fuzzy set $A$ on a set $X$ is characterized by a mapping $m: X \to [0, 1]$, called the membership function.

A fuzzy set is denoted as $A = (X, m)$. A fuzzy graph [19] $\xi = (V, \sigma, \mu)$ is a non-empty set V together with a pair of functions $\sigma: V \to [0, 1]$ and $\mu: V \times V \to [0, 1]$ such that for all $u, v \in V$, $\mu(u,v) \leq \sigma(u) \wedge \sigma(v)$ (here $x \wedge y$ denotes the minimum of $x$ and $y$). Partial fuzzy subgraph $\xi' = (V, \tau, v)$ of $\xi$ is such that $\tau(v) \leq \sigma(v)$ for all $v \in V$ and $\mu(u,v) \leq V(u,v)$ for all $u,v \in V$. Fuzzy subgraph [12] $\xi'' = (P, \sigma', \mu')$ of $\xi$ is such that $P \subseteq V$, $\sigma'(u) = \sigma(u)$ for all $u \in P$, $\mu'(u,v) = \mu(u,v)$ for all $u,v \in P$.

A fuzzy graph is complete [13] if $\mu(u,v) = \sigma(u) \wedge \sigma(v)$ for all $u,v \in V$. The degree of vertex $u$ is $d(u) = \sum_{(u,v) \in \xi} \mu(u,v)$. The minimum degree of $\xi$ is $\delta(\xi) = \wedge \{d(u) | u \in V\}$. The maximum degree of $\xi$ is $\Delta(\xi) = \vee \{d(u) | u \in V\}$. The total degree [13] of a vertex $u \in V$ is $td(u) = d(u) + \sigma(u)$. A fuzzy graph $\xi = (V, \sigma, \mu)$ is said to be regular [13] if $d(v) = k$, a positive real number, for all $v \in V$. If each vertex of $\xi$ has same total degree k, then $\xi$ is said to be a totally regular fuzzy graph. A fuzzy graph is said to be irregular [16], if there is a vertex which is adjacent to vertices with distinct degrees. A fuzzy graph is said to be neighbourly irregular [16], if every two adjacent vertices of the graph have different degrees.

A fuzzy graph is said to be totally irregular, if there is a vertex which is adjacent to vertices with distinct total degrees. If every two adjacent vertices have distinct total degrees of a fuzzy graph then it is called neighbourly total irregular [16]. A fuzzy graph is called highly irregular [16] if every vertex of $G$ is adjacent to vertices with distinct degrees. The complement [12] of fuzzy graph $\xi = (V, \sigma, \mu)$ is the fuzzy graph $\bar{\xi} = (V, \sigma', \mu')$ where $\sigma'(u) = \sigma(u)$ for all $u \in V$ and

$$\mu'(u,v) = \begin{cases} 0 & \text{if } \mu(u,v) > 0, \\ \sigma(u) \wedge \sigma(v), & \text{otherwise.} \end{cases}$$

Let $X$ be a nonempty set. An interval-valued fuzzy set $A$ in $V$ is defined by $A = \{(x, [\mu_{A^-(x)}, \mu_{A^+(x)}]) : x \in V\}$, where $\mu_{A^-(x)}$ and $\mu_{A^+(x)}$ are fuzzy subsets of $V$ such that $\mu_{A^-(x)} \leq \mu_{A^+(x)}$ for all $x \in V$.

For any two interval-valued fuzzy sets $A = [\mu_{A^-(x)}, \mu_{A^+(x)}]$ and $B = [\mu_{B^-(x)}, \mu_{B^+(x)}]$ in $V$ we have:

- $A \cup B = \{(x, \max(\mu_{A^-}(x), \mu_{B^-}(x)), \max(\mu_{A^+}(x), \mu_{B^+}(x))) : x \in V\}$,
- $A \cap B = \{(x, \min(\mu_{A^-}(x), \mu_{B^-}(x)), \min(\mu_{A^+}(x), \mu_{B^+}(x))) : x \in V\}$.

If $G^* = (V, E)$ is a graph, then by an interval-valued fuzzy relation B on a set E we mean an interval - valued fuzzy set such that.

$$\mu_{B^-}(xy) \leq \min\left(\mu_{A^-}(x), \mu_{A^-}(y)\right),$$
$$\mu_{B^+}(xy) \leq \min\left(\mu_{A^+}(x), \mu_{A^+}(y)\right) \text{ for all } xy \in E.$$

By an interval - valued fuzzy graph of a graph $G^* = (V, E)$ we mean a pair $G = (A, B)$, where $A = [\mu_{A^-}, \mu_{A^+}]$ is an interval-valued fuzzy set on V and $B = [\mu_{B^-}, \mu_{B^+}]$ is an interval-valued fuzzy relation on $E$.

The graph $G$ is called complete interval-valued fuzzy graph if

$$\mu_{B^+}(xy) = \min\left(\mu_{A^+}(x), \mu_{A^+}(y)\right) \text{ and } \mu_{B^-}(xy) = \min\left(\mu_{A^-}(x), \mu_{A^-}(y)\right)$$

for all $x, y \in V$.

An interval-valued fuzzy graph $G = (A, B)$ of a given graph $G^* = (V, E)$ is called an interval-valued strong fuzzy graph if

$$\mu_{B^-}(xy) = \min\left(\mu_{A^-}(x), \mu_{A^-}(y)\right) \text{ and } \mu_{B^+}(xy) = \min\left(\mu_{A^+}(x), \mu_{A^+}(y)\right)$$

for all $xy \in E$.

The complement of a strong interval-valued fuzzy graph $G$ is $\bar{G} = (\bar{A}, \bar{B})$ where $\bar{A} = [\bar{\mu}_{A^-}(x), \bar{\mu}_{A^+}(x)]$ is an interval-valued fuzzy set on $\bar{V}$ and $\bar{B} = [\bar{\mu}_{B^-}, \bar{\mu}_{B^+}]$ is an interval-valued fuzzy set on $\bar{E} \subseteq \bar{V} \times \bar{V}$ such that

(1) $\bar{V} = V$,

(2) $\bar{\mu}_{A^-}(x) = \mu_{A^-}(x)$ and $\bar{\mu}_{A^+}(x) = \mu_{A^+}(x)$ for all $x \in V$,

Irregular Interval–Valued Fuzzy Graphs

(3) $\bar{\mu}_{B^-}(xy) = \begin{cases} 0 & \text{If } \mu_{B^-}(xy) > 0, \\ \mu_{A^-}(x) \wedge \mu_{A^-}(y) & \text{If } \mu_{B^-}(xy) = 0. \end{cases}$

(4) $\bar{\mu}_{B^+}(xy) = \begin{cases} 0 & \text{If } \mu_{B^+}(xy) > 0, \\ \mu_{A^+}(x) \wedge \mu_{A^+}(y) & \text{If } \mu_{B^+}(xy) = 0. \end{cases}$

**Definition 1[1].** Let $G = (A, B)$ be an interval-valued fuzzy graph where $A = [\mu_{A^-}, \mu_{A^+}]$ and $B = [\mu_{B^-}, \mu_{B^+}]$ be two interval-valued fuzzy sets on a non-empty finite set $V$ and $E \subseteq V \times V$ respectively. The graph $G$ is called complete interval-valued fuzzy graph if $\mu_{B^-}(xy) = \min\left(\mu_{A^-}(x), \mu_{A^-}(y)\right)$ and
$$\mu_{B^+}(xy) = \min\left(\mu_{A^+}(x), \mu_{A^+}(y)\right) \text{ for all } x, y \in V.$$

**Definition 2.** Let $G = (A, B)$ be an interval-valued fuzzy graph where $A = [\mu_{A^-}, \mu_{A^+}]$ and $B = [\mu_{B^-}, \mu_{B^+}]$ be two interval - valued fuzzy sets on a non- empty finite set V and $E \subseteq V \times V$ respectively. The total degree of a vertex $u \in V$ is denoted by $td(u)$ and defined as $td(u) = [td^+(u), td^-(u)]$ where
$$td^+(u) = \sum_{uv \in E} \mu_{B^+}(uv) + \mu_{A^+}(u), \quad td^-(u) = \sum_{uv \in E} \mu_{B^-}(uv) + \mu_{A^-}(u).$$

If the total degrees of all vertices of an interval-valued fuzzy graph are equal, then the graph is said to be totally regular interval-valued fuzzy graph.

## 3. Some definitions related to interval-valued fuzzy graphs

The degree of a vertex of an interval-valued fuzzy graph is defined below.

**Definition 3.** Let $G = (A, B)$ be an interval-valued fuzzy graph where $A = [\mu_{A^-}, \mu_{A^+}]$ and $B = [\mu_{B^-}, \mu_{B^+}]$ be two interval-valued fuzzy sets on a non-empty finite set V and $E \subseteq V \times V$ respectively. The positive degree of a vertex $u \in G$ is $d^+(u) = \sum_{uv \in E} \mu_{B^+}(uv)$. Similarly, negative degree of a vertex $u \in G$ is $d^-(u) = \sum_{uv \in E} \mu_{B^-}(uv)$. The degree of a vertex $u$ is $d(u) = [d^-(u), d^+(u)]$.

If $d^+(u) = k_1, d^-(u) = k_2$ for all $u \in V$, $k_1, k_2$ are two real numbers, then the graph is called $[k_1, k_2]$-regular interval valued fuzzy graph.

**Example 1.** We consider an interval-valued fuzzy graph show here. We have $d^-(x) = 0.1 + 0.1 = 0.2$, $d^+(x) = 0.3 + 0.4 = 0.7$. So, $d(x) = (0.2, 0.7)$. Similarly, $d(y) = (0.3, 0.7)$ and $d(z) = (0.3, 0.8)$.

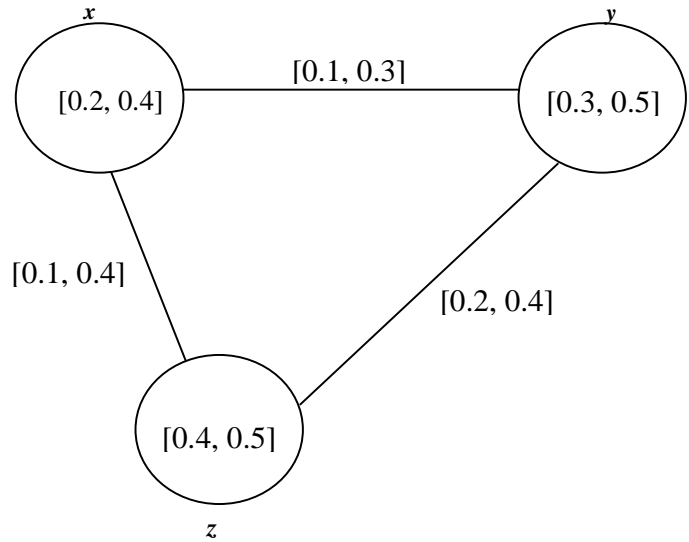

**Figure 1:** An example of interval-valued fuzzy graph

The order and size of an interval-valued fuzzy graph are important terms. They are defined below.

**Definition 4.** Let $G = (A, B)$ be an interval-valued fuzzy graph where $A = [\mu_{A^-}, \mu_{A^+}]$ and $B = [\mu_{B^-}, \mu_{B^+}]$ be two interval-valued fuzzy sets on a non-empty finite set V and $E \subseteq V \times V$ respectively. The order of G is denoted by O(G) and is defined by $O(G) = [O^-(G), O^+(G)]$ where $O^-(G) = \sum_{u \in V} \mu_{A^-}(u)$ and $O^+(G) = \sum_{u \in V} \mu_{A^+}(u)$.

**Definition 5.** Let $G = (A, B)$ be an interval-valued fuzzy graph where $A = [\mu_{A^-}, \mu_{A^+}]$ and $B = [\mu_{B^-}, \mu_{B^+}]$ be two interval-valued fuzzy sets on a non-empty finite set V and E respectively. The size of G is defined by $S(G) = [S^-(G), S^+(G)]$ where $S^-(G) = \sum_{\substack{uv \in E \\ u \neq v}} \mu_{B^-}(uv)$ and $S^+(G) = \sum_{\substack{uv \in E \\ u \neq v}} \mu_{B^+}(uv)$.

**Example 2.** For the interval-valued fuzzy graph of Figure 1, $O(G) = (0.9, 1.4)$ and $S(G) = (0.4, 1.1)$.

Irregular Interval–Valued Fuzzy Graphs

**Definition 6.** Let $G = (A, B)$ be an interval-valued fuzzy graph where $A = [\mu_{A^-}, \mu_{A^+}]$ and $B = [\mu_{B^-}, \mu_{B^+}]$ be two interval-valued fuzzy sets on a non-empty finite set $V$ and $E \subseteq V \times V$ respectively. The underlying crisp graph of G is the crisp graph $G' = (V', E')$ where $V' = \{v \mid \mu_{A^+}(v) > 0 \text{ and } \mu_{A^-}(v) > 0\}$ and $E' = \{(u,v) \mid \mu_{B^+}(uv) > 0 \text{ and } \mu_{B^-}(uv) > 0\}$.

**Definition 7.** An interval-valued fuzzy graph is said to be connected if it's underlying crisp graph is connected.

**Theorem 1.** Let $G$ be a regular interval-valued fuzzy graph where induced crisp graph $G'$ is an even cycle. Then $G$ is regular interval-valued fuzzy graph if and only if either $\mu_{B^+}$ or $\mu_{B^-}$ is constant functions or alternate edges have same positive membership values and negative membership values.

**Proof.** Let $G = (A, B)$ be a regular interval-valued fuzzy graph where $A = [\mu_{A^-}, \mu_{A^+}]$ and $B = [\mu_{B^-}, \mu_{B^+}]$ be two interval-valued fuzzy sets on a non-empty finite set $V$ and $E \subseteq V \times V$ respectively and underlying crisp graph $G'$ of $G$ be an even cycle. If either $\mu_{B^-}$ or $\mu_{B^+}$ is constant function or alternate edges have same positive and negative membership values, then $G$ is a regular interval-valued fuzzy graph. Conversely, suppose G is a $[k_1, k_2]$-regular interval-valued fuzzy graph. Let $e_1, e_2, \ldots e_n$ be the edges of $G'$ in order. As

$$\mu_{B^+}(e_i) = \begin{cases} c_1 & \text{if i is odd,} \\ k_1 - c_1 & \text{if i is even.} \end{cases}$$

$$\mu_{B^-}(e_i) = \begin{cases} c_2 & \text{if i is odd,} \\ k_2 - c_2 & \text{if i is even.} \end{cases}$$

If $c_1 = k_1 - c_1$, then $\mu_{B^+}$ is constant. If $c_1 \neq k_1 - c_1$, then alternate edges have same positive and negative membership values. Similarly, for $\mu_{B^-}$. Hence the result. □

**Theorem 2.** The size of a $(k_1, k_2)$-regular interval-valued fuzzy graph is $(\frac{Pk_1}{2}, \frac{Pk_2}{2})$ where $P = |V|$.

**Proof.** Let $G = (A, B)$ be an interval-valued fuzzy graph where $A = [\mu_{A^-}, \mu_{A^+}]$ and $B = [\mu_{B^-}, \mu_{B^+}]$ be two interval-valued fuzzy sets on a non-empty finite set $V$ and $E \subseteq V \times V$ respectively. The size of G is $S(G) = \left[ \sum_{u \neq v} \mu_{B^-}(uv), \sum_{u \neq v} \mu_{B^+}(uv) \right]$.

Now, $\sum_{v \in V} d(v) = 2[\sum_{uv \in E} \mu_{B^+}(uv), \sum_{uv \in E} \mu_{B^-}(uv)] = 2 S(G)$.

Thus, $2S(G) = \sum_{v \in V} d(v)$, i.e. $2S(G) = [\sum_{v \in V} k_1, \sum_{v \in V} k_2]$.

This gives $2S(G) = [P k_1, P k_2]$. Hence the result. □

**Theorem 3.** If G is $[k, k']$-totally regular interval-valued fuzzy graph, then $2S(G) + O(G) = [P k, Pk']$ where $P = |V|$.

**Proof.** Let $G = (A, B)$ be an interval-valued fuzzy graph where $\sum_{u \neq v} \mu_{B^-}(uv)$, $\sum_{u \neq v} \mu_{B^+}(uv)$ be two interval-valued fuzzy sets on a non-empty finite set $V$ and $V \times V$ respectively. Since G is a $[k, k']$-totally regular interval-valued fuzzy graph. So $k = td^+(v) = d^+(v) + \mu_{A^+}(v)$ and $k' = td^-(v) = d^-(v) + \mu_{A^-}(v)$ for all $v \in V$.
Therefore
$$\sum_{v \in V} k = \sum_{v \in V} d^+(v) + \sum_{v \in V} \mu_{A^+}(v) \text{ and } \sum_{v \in V} k' = \sum_{v \in V} d^-(v) + \sum_{v \in V} \mu_{A^-}(v).$$
$Pk = 2S^+(G)$ and $Pk' = 2S^-(G)$. So
$Pk + Pk' = 2(S^+(G) + S^-(G)) + O^+(G) + O^-(G)$.
Hence $2S(G) + O(G) = [Pk, Pk']$. □

### 4. Irregular interval-valued fuzzy graphs
Irregular interval-valued fuzzy graphs are important as regular interval-valued fuzzy graphs. We now define it.

**Definition 8.** Let $G = (A, B)$ be an interval-valued fuzzy graph where $A = [\mu_{A^-}, \mu_{A^+}]$ and $B = [\mu_{B^-}, \mu_{B^+}]$ be two interval-valued fuzzy sets on a non-empty finite set $V$ and $E \subseteq V \times V$ respectively. $G$ is said to be irregular interval-valued fuzzy graph if there exists a vertex which is adjacent to a vertex with distinct degrees.

**Example 3.** Let $G = (A, B)$ be an interval - valued fuzzy graph where $A = [\mu_{A^-}, \mu_{A^+}]$ and $B = [\mu_{B^-}, \mu_{B^+}]$ be two interval-valued fuzzy sets on a non-empty finite set $V$ and $E \subseteq V \times V$ respectively, where $V = \{v_1, v_2, v_3, v_4\}$.
$d(v_1) = [0.8, 1]$, $d(v_2) = [0.8, 1]$, $d(v_3) = [1.2, 1.4]$, $d(v_4) = [0.4, 0]$.
Here $d(V_2) \neq d(V_3)$. So this graph is an example of irregular interval-valued fuzzy graph, show in Figure 2.

Neighbourly irregular interval-valued fuzzy graph is a special case of irregular interval-valued fuzzy graph.

Irregular Interval–Valued Fuzzy Graphs

**Definition 9.** Let $G$ be a connected interval-valued fuzzy graph. Then $G$ is called neighbourly irregular interval-valued fuzzy graph if for every two adjacent vertices of $G$ have distinct degrees.

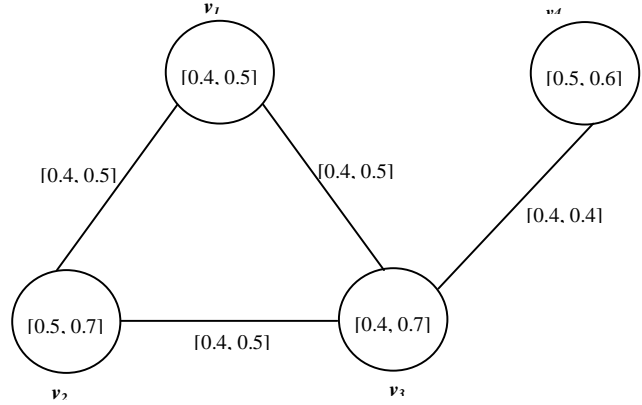

**Figure 2:** An example of irregular interval-valued fuzzy graph.

**Definition 10.** Let $G = (A, B)$ be an interval-valued fuzzy graph where $A = [\mu_{A^-}, \mu_{A^+}]$ and $B = [\mu_{B^-}, \mu_{B^+}]$ be two interval-valued fuzzy sets on a non-empty finite set $V$ and $E \subseteq V \times V$ respectively. $G$ is said to be totally irregular interval-valued fuzzy graph if there exists a vertex which is adjacent to a vertex with distinct total degrees.

**Definition 11.** Let $G$ be a connected interval-valued fuzzy graph. Then $G$ is called highly irregular interval-valued fuzzy graph if every vertex of $G$ is adjacent to vertices with distinct degrees.

**Example 4.** Let $G = (A, B)$ be an interval-valued fuzzy graph where $A = [\mu_{A^-}, \mu_{A^+}]$ and $B = [\mu_{B^-}, \mu_{B^+}]$ be two interval-valued fuzzy sets on a non-empty finite set $V$ and $E \subseteq V \times V$ respectively, where $V = \{v_1, v_2, v_3, v_4\}$. $v_1 = [0.4, 0.5]$, $v_2 = [0.5, 0.6]$, $v_3 = [0.4, 0.5], v_4 = [0.4, 0.4], \mu_{B^-}(v_1 v_2) = 0.3, \mu_{B^+}(v_1 v_2) = 0.4, \mu_{B^-}(v_2 v_3) = 0.2,$ $\mu_{B^+}(v_2 v_3) = 0.4, \mu_{B^-}(v_3 v_4) = 0.3, \mu_{B^+}(v_3 v_4) = 0.4, \mu_{B^-}(v_2 v_4) = 0.2,$ $\mu_{B^+}(v_2 v_4) = 0.4$.
So we have $d(v_1) = [0.3, 0.4], d(v_2) = [0.8, 1.2], d(v_3) = [0.5, 0.8],$
$d(v_4) = [0.5, 0.8])$. Here the interval-valued fuzzy graph is highly irregular but not neighbourly irregular as $d(v_3) = d(v_4)$.

**Theorem 4.** Let G be an interval-valued fuzzy graph. Then G is highly irregular interval-valued fuzzy graph and neighbourly irregular interval-valued fuzzy graph if and only if the degrees of all vertices of *G* are distinct.

**Proof.** Let $G = (A, B)$ be an interval-valued fuzzy graph where $A = [\mu_{A^-}, \mu_{A^+}]$ and $B = [\mu_{B^-}, \mu_{B^+}]$ be two interval-valued fuzzy sets on a non-empty finite set *V* and $V \times V$ respectively. Let $V = \{v_1, v_2, ..., v_n\}$. We assume that *G* is highly irregular and neighbourly irregular interval-valued fuzzy graphs. Let the adjacent vertices of $u_1$ be $u_2, u_3, ..., u_n$ with degrees $[k_2^-, k_2^+], [k_3^-, k_3^+], ..., [k_n^-, k_n^+]$ respectively. As *G* is highly and neighbourly irregular, $d(u_1) \neq d(u_2) \neq d(u_3) \neq ... \neq d(u_n)$. So it is obvious that all vertices are of distinct degrees.

Conversely, assume that the degrees of all vertices of *G* are distinct. This means that every two adjacent vertices have distinct degrees and to every vertex the adjacent vertices have distinct degrees. Hence, *G* is neighbourly irregular and highly irregular interval-valued fuzzy graphs.

**Theorem 5.** Let G be an interval-valued fuzzy graph. If G is neighbourly irregular and $\mu_{A^-}$, $\mu_{A^+}$ are constant functions, then G is a neighbourly total irregular interval-valued fuzzy graph.

**Proof.** Let $G = (A, B)$ be an interval-valued fuzzy graph where $A = [\mu_{A^-}, \mu_{A^+}]$ and $B = [\mu_{B^-}, \mu_{B^+}]$ be two interval-valued fuzzy sets on a non-empty finite set *V* and $E \subseteq V \times V$ respectively.

Assume that *G* is a neighbourly irregular interval-valued fuzzy graph, i.e. the degrees of every two adjacent vertices are distinct. Consider two adjacent vertices $u_1$ and $u_2$ with distinct degrees $[k_1^-, k_1^+]$ and $[k_2^-, k_2^+]$ respectively. Also let, $\mu_{A^-}(u) = c_1$, $\mu_{A^+}(u) = c_2$ for all $u \in V$ where $c_1, c_2 \in [0, 1]$ are constants. Therefore,
$td(u_1) = [d^-(u_1) + c_1, d^+(u_1) + c_2] = [k_1^- + c_1, k_1^+ + c_2]$
$td(u_2) = [d^-(u_2) + c_1, d^+(u_2) + c_2] = [k_2^- + c_1, k_2^+ + c_2]$. Clearly, $td(u_1) \neq td(u_2)$. Therefore for any two adjacent vertices $u_1$ and $u_2$ with distinct degrees, it's total degrees are also distinct, provided $\mu_{A^-}$, $\mu_{A^+}$ are constant functions. The above argument is true for every pair of adjacent vertices in *G*.

## 5. Conclusions

Graph theory is an extremely useful tool in solving the combinatorial problems in different areas including geometry, algebra, number theory, topology, operations research and computer science. In this paper, we have described degree of a vertex, order, size and underlying crisp graph of an interval-valued fuzzy graph. The necessary and sufficient conditions for an interval-valued fuzzy graph to be the regular interval-valued fuzzy graphs have been presented. Size of an interval-valued fuzzy graph a relation between size and order of an interval-valued fuzzy graph have been calculated. We have defined irregular interval-valued fuzzy graphs, neighbourly irregular, totally and highly irregular interval-valued fuzzy graphs. Some relations about the defined graphs have been proved.